# Superconductor terahertz metamaterial

Jianqiang Gu,[1,2] Ranjan Singh,[1] Zhen Tian,[1,2] Wei Cao,[1] Qirong Xing,[2] Jiaguang Han,[2,3] and Weili Zhang[1,*]

[1]*School of Electrical and Computer Engineering, Oklahoma State University, Stillwater, Oklahoma 74078, USA*
[2]*Center for Terahertz waves and College of Precision Instrument and Optoelectronics Engineering, Tianjin University, and the Key Laboratory of Optoelectronics Information and Technical Science (Ministry of Education), Tianjin 300072, P. R. China*
[3]*Department of Physics, National University of Singapore, 2 Science Drive 3, Singapore*
*\* weili.zhang@okstate.edu*

**Abstract:** We characterize the behaviour of split ring resonators made up of high-transition temperature YBCO superconductor using terahertz time-domain spectroscopy. The superconductor metamaterial shows sharp change in the transmission spectrum at the fundamental inductive-capacitive resonance and the dipole resonance as the temperature dips below the transition temperature. Our results reveal that the high performance of such a metamaterial is limited by material imperfections and defects such as cracks, voids and secondary phases which play dominant role in partially impeding the flow of current causing dissipation of energy and electrical resistance to appear in the superconductor film.

## References and links

1. J. Pendry, "Negative Refraction Makes a Perfect Lens," Phys. Rev. Lett. **85**, 3966 (2000).
2. N. Fang, H. Lee, C. Sun, and X. Zhang, "Sub–Diffraction-Limited Optical Imaging with a Silver Superlens," Science **308**, 534 (2005).
3. J. O'Hara, R. Singh, I. Brener, E. Smirnova, J. Han, A. Tylor, and W. Zhang, "Thin-film sensing with planar terahertz metamaterials: sensitivity and limitations," Opt. Express **16**, 1786 (2008).
4. M. Nagel, P. Haring-Bolivar, M. Brucherseifer, H. Kurz, A. Bosserhoff, and R. Buttner, "Integrated planar terahertz resonators for femtomolar sensitivity label-free detection of DNA hybridization," Appl. Opt. **41**, 2074 (2002).
5. C. Debus and P. Haring Bolivar, "Frequency selective surfaces for high sensitivity terahertz sensing," Appl. Phys.Lett. **91**, 184102 (2007).
6. I. A. Naib, C. Jansen, and M. Koch, "Thin film sensing with planar asymmetric metamaterial resonators," Appl. Phys. Lett. **93**, 083507 (2008).
7. A. Ishimaru, S. Jaruwatanadilok, and Y. Kuga, "Generalized surface plasmon resonance sensors using metamaterials and negative index materials," Progress In Electromagnetics Research **51**, 139 (2005).
8. D. Schurig, J. Mock, B. Justice, S. Cummer, J. Pendry, A. Starr, and D. Smith, "Metamaterial Electromagnetic Cloak at Microwave Frequencies," Science **314**, 977 (2006).
9. R. Liu, C. Ji, J. Mock, J. Chin, T. Cui, and D. R. Smith, "Broadband Ground-Plane Cloak," Science **323**, 366 (2009).
10. B. Edwards, A. Alù, M. Silveirinha, and N. Engheta, "Experimental Verification of Plasmonic Cloaking at Microwave Frequencies with Metamaterials," Phys. Rev. Lett. **103**, 153901 (2009).
11. S. Zhang, Y. Park, J. Li, X. Lu, W. Zhang, and X. Zhang, "Negative Refractive Index in Chiral Metamaterials," Phys. Rev. Lett. **102**, 023901 (2009).
12. A. Papakostas, A. Potts, D. Bagnall1, S. Prosvirnin, H. Coles, and N. Zheludev, "Optical Manifestations of Planar Chirality," Phys. Rev. Lett. **90**, 107404 (2003).
13. R. Singh, E. Plum, C. Menzel, C. Rockstuhl, A. Azad, R. Cheville, F. Lederer, W. Zhang, and N. Zheludev, "Terahertz metamaterial with asymmetric transmission," Phys. Rev. B **80**, 153104 (2009).
14. N. Liu, L. Langguth, T. Weiss, J. Kästel, M. Fleischhauer, T. Pfau, and H. Giessen, "Plasmonic analogue of electromagnetically induced transparency at the Drude damping limit," Nature Materials **8**, 758 (2009).
15. R. Singh, C. Rockstuhl, F. Lederer, and W. Zhang, "Coupling between a dark and a bright eigenmode in a terahertz metamaterial," Phys. Rev. B **79**, 085111 (2009).
16. S. Chiam, R. Singh, C. Rockstuhl, F. Lederer, W. Zhang, and A. Bettiol, "Analogue of electromagnetically induced transparency in a terahertz metamaterial," Phys. Rev. B **80**, 153103 (2009).

17. H. Chen, W. Padilla, J. Zide, A. Gossard, A. Taylor, and R. Averitt, "Active terahertz metamaterial devices," Nature **444,** 597 (2006).
18. H. Chen, W. Padilla, M. Cich, A. Azad, R. Averitt, and A. Taylor, "A metamaterial solid-state terahertz phase modulator," Nature Photonics **3**, 148 (2009).
19. O. Paul, C. Imhof, B. Lägel, S. Wolff, J. Heinrich, S. Höfling, A. Forchel, R. Zengerle, R. Beigang, and M. Rahm, "Polarization-independent active metamaterial for high-frequency terahertz modulation," Opt. Express **17**, 819 (2009).
20. C. Soukoulis, J. Zhou, T. Koschny, M. Kafesaki, and E. Economou, "The Science of Negative Index Materials," J. Phys. : Condens. Matter. **20**, 304217 (2008).
21. G. Dolling, M.Wegener, C. Soukoulis, and S. Linden, "Negative-index metamaterial at 780 nm wavelength," Opt. Lett. **32**, 53 (2007).
22. U. Chettiar, A. Kildishev, H. Yuan, W. Cai, S. Xiao, V. Drachev, and V. Shalaev, "Dual-band negative index metamaterial: double negative at 813 nm and single negative at 772 nm," Opt. Lett. **32**, 1671 (2007).
23. J. Bednorz, and K. Müller "Possible high $T_c$ superconductivity in the Ba−La−Cu−O system," Z. Phys. B **64**, 189 (1986).
24. M. Ricci, N. Orloff, and S. Anlage, "Superconducting metamaterials," Appl. Phys. Lett. **87**, 034102 (2005).
25. V. Fedotov, J. Shi1, A. Tsiatmas, P. Groot, Y. Chen, and N. Zheludev, "Fano Resonances in High-Tc Superconducting Metamaterial," arXiv:1001.4154.
26. S. Speller, R. Dinner, and C. Grovenor, "Growth of $Tl_2Ba_2CaCu_2O_8$ superconducting thin films on curved substrates for metamaterials applications," Journal of Crystal Growth **310**, 4081 (2008).
27. D. Grischkowsky, S. Keiding, M. Exter, and Ch. Fattinger, "Far infrared time domain spectroscopy with terahertz beams of dielectrics and semiconductors," J. Opt. Soc. Am. B **7,** 2006 (1990).
28. R. Singh, C. Rockstuhl, F. Lederer, and W. Zhang, "The impact of nearest neighbor interaction on the resonances in terahertz metamaterials," Appl. Phys. Lett. **94**, 064102 (2009).
29. R. Singh, Z. Tian, J. Han, C. Rockstuhl, J. Gu, and W. Zhang, "Cryogenic temperatures as a path toward high-Q terahertz metamaterials," App. Phys. Lett. **96**, 071114 (2010).
30. A. Azad, J. Dai, and W. Zhang, "Transmission properties of terahertz pulses through subwavelength double split ring resonators," Opt. Lett. **31**, 634 (2006).
31. W. Padilla, A. Taylor, C. Highstrete, M. Lee, and R. Averitt, "Dynamical electric and magnetic metamaterial response at terahertz frequencies," Phys. Rev. Lett. **96**, 107401 (2006).
32. F. London, and H. London, "The Electromagnetic Equations of the Supraconductor," Proc. Roy. Soc. **A149**, 71.
33. M. Khazan, "Time-domain terahertz spectroscopy and its application to the study of high-$T_c$ superconductor thin films," Ph.D Thesis, University Hamburg (2002).
34. I. Wilke, M. Khazan, C. Rieck, P. Kuzel, T. Kaiser, C. Jaekel, and H. Kurz, "Investigation of Terahertz Emission Phenomena of High-$T_c$ Superconductors," J. Appl. Phys. **87**, 2984 (2000).
35. R. Singh, E. Smirnova, A. Taylor, J. O'Hara, and W. Zhang, "Optically thin terahertz metamaterials," Opt. Express **16**, 6537 (2008).
36. R. Singh, A. Azad, J. O'Hara, A. Taylor, and W. Zhang, "Effect of metal permittivity on resonant properties of terahertz metamaterials," Opt. Lett. **33**, 1506 (2008).

## 1. Introduction

The emerging field of metamaterial (MM) has opened gateway to unprecedented electromagnetic properties and functionality unattainable from naturally occurring materials, thus enabling a family of MM based devices such as superlens [1,2], sensors [3-7], cloaks [8-10], chiral devices [11-13], electromagnetically-induced-transparency (EIT) components for slow light [14-16] and modulators [17-19]. However, in spite of these and other advances, progress toward practical implementation, particularly at higher frequencies, has been hampered by high absorption losses [20-22]. A large majority of the existing MM designs rely on the use of metallic structure sitting on dielectric substrates. As the frequency of operation is pushed higher towards the terahertz, infrared and visible, the ohmic losses quickly render the current MM approaches impractical. Thus, a top priority is to reduce the absorption losses to levels suitable for device applications. This would require MM designs that do not depend solely on metallic structures. One approach would be to utilize sub-wavelength arrays of split ring resonator (SRR) structures made up of zero resistance superconductors which allow dissipation less flow of electrical current.

Over twenty years back the discovery of high transition temperature (high-$T_c$) superconductors (HTS) [23] promised to conduct electric current without resistance at liquid nitrogen temperature still presents tremendous challenge to our understanding. All known

HTS are cuprate (copper-oxide) materials and they have great potential to find applications in the design of low loss MMs. M. Ricci and co-workers fabricated yttrium barium copper oxide (YBCO) MM and studied its tunability under magnetic field control at microwave frequencies [24]. Fedotov *et al*. showed the Fano resonances in microwave YBCO MM and observed the interaction between electrical and magnetic dipoles [25]. In 2007, another type of HTS, thallium barium calcium copper oxide (TBCCO) was used by S. Speller and co-workers to fabricate MM structures on curved surfaces [26]. Thus, most of the previous works on superconductor MMs were focused in the microwave regime and since the behavior of HTS is frequency dependent, it becomes highly essential to investigate HTS MMs at higher frequencies of operation.

In this work, we characterize the response of YBCO MM at terahertz frequencies using liquid nitrogen cryostat installed in a terahertz time-domain spectroscopy (THz-TDS) system [27]. Interestingly, the real part of the conductivity of YBCO in the terahertz regime lies between that of a dielectric and metal and is on the order of $10^5 \sim 10^6$ while its temperature is varied from room temperature to the liquid nitrogen temperature. Using simple liquid nitrogen cryogenics, the transition from normal state to superconducting state can be achieved in YBCO HTS which makes it really attractive for the development of temperature tunable terahertz devices. We measured the response of YBCO single SRRs fabricated on sapphire substrates at different temperatures under normal incidence with the terahertz electric field oriented along the gap arm of the SRR. The modulation of LC and dipole resonances is clearly observed as the temperature approaches the transition value of the superconductor. Simplified conductivity model plus the two-fluid thesis are applied to analyze the changes in transmission resonances and based on this assumption the full-wave CST simulations could reproduce the experimental observations and gave further insight about the behavior of superconductor MM at temperatures below the transition point which we could not access experimentally.

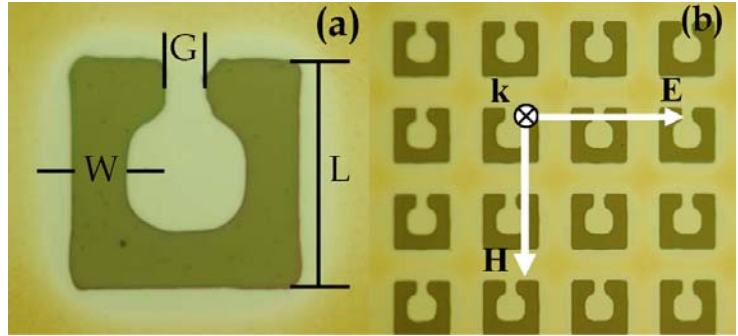

Fig. 1. Microcopic image of (a) YBCO MM unit cell with structural parameters, $W$ = 8 μm, $G$ = 5 μm, $L$ = 32 μm (b) MM sample array with periodicity of the unit cells is $P$ = 52 μm. The incident E field is polarized parallel to the gap of the single SRRs.

## 2. Experimental

The sample is made from a commercial 280 nm-thick YBCO film which typically has 86 K as the transition temperature and 2.3 mA maximum current grown on a 500 μm thick sapphire substrate (THEVA, Germany). By using conventional photolithographic exposure we patterned a 3-μm-thick negative photoresist, NR7-3000P film into single SRR shape on the YBCO film as the protective layer. Then the sample is wet etched in 0.1% nitric acid for 90 s to remove YBCO from other parts on the wafer that did not have the photoresist protection followed by lift-off in pure acetone. The sample array size is 5 mm × 5 mm with SRR periodicity $P$ = 52 μm, and the structural dimensions of unit cells are length $L$ = 32 μm, width

$W = 8$ μm and gap $G = 5$ μm, as shown in the microscopic image in Fig. 1 [28]. The sample was measured on a standard cryogenic THz-TDS system which has a bandwidth from 0.3 to 2.5 THz with signal to noise ratio as high as 2000:1 [29]. The terahertz beam is focused to a 2.0 mm diameter waist to ensure a measurement without frequency filtering and the beam excites several thousand SRRs. The sample is set with the SRR gap along the incident **E** field in order to excite the fundamental LC mode resonance and the dipole resonance. We focus more on the LC mode since at this resonance the SRRs provide a strong magnetic response with negative permeability, required for achieving negative index of refraction. A bare sapphire substrate is used as the reference for all sets of measurements. At a particular temperature the transmitted sub-picosecond terahertz pulse through the sample and the reference is measured in time domain thrice and averaged. The transmission in frequency spectrum is defined as $|E_s(\omega)/E_r(\omega)|$, where $E_s(\omega)$ and $E_r(\omega)$ are the Fourier transform of the averaged time-domain sample and reference signals, respectively. We wish to measure the frequency dependent amplitude and phase of the YBCO MM at reduced temperatures.

By filling the vacuum chamber with liquid nitrogen, the temperature of the sample starts to go down from room temperature (297 K) to 85 K. The sample and the reference were measured at 297, 200, 150, 100 and 85 K. The measured time domain pulses as shown in Fig. 2a reveals the dramatic change of the pulse shape upon cooling the sample from room temperature to near the superconducting transition temperature at 85 K. The reshaping of the pulse is direct consequence of superconductor's kinetic inductance and this inductance is entirely due to the superconducting carrier's kinetics since a pure superconductor acts as an ideal inductor-the impedance of which is zero for dc current and imaginary for ac current, however no power is dissipated through imaginary impedance. Figure 2b shows the amplitude transmission spectrum at different temperatures. At room temperature, there are two weak re-sonance dips lying around 0.7 and 1.7 THz, the modulation depth only reaching 0.78 and 0.75 respectively due to the weak conductivity of YBCO. As the temperature is reduced, the resonance dips become more and more pronounced. The phase modulation controlled by temperature change can be seen in Fig. 2c. According to Kramers-Kronig relation, at 0.7 and 1.7 THz where the amplitude transmission have minima, the change of the phase is maximum and vice versa. Figure 2d shows the percentage change in amplitude transmission at the LC and dipole resonance features with reducing temperatures. A noteworthy feature is that the sharpest change occurs when the temperature of YBCO MM transitions from 100 to 85 K, thus the transmission at the LC resonance sharply dropping from 0.68 to 0.58.

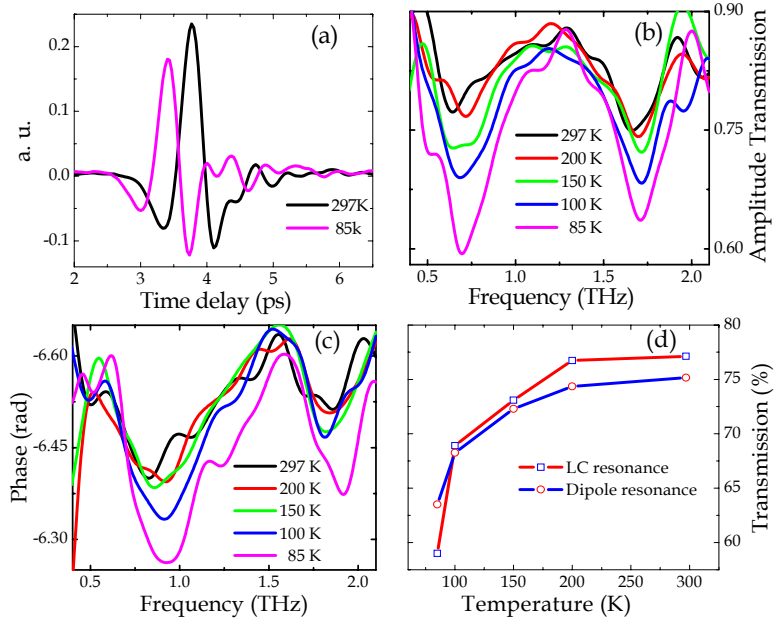

Fig. 2. (color online). Measured (a) sub-picosecond transmitted pulses through the MM at 297K and 85K, (b) amplitude transmission spectrum, (c) phase change spectrum, and (d) percentage amplitude transmission at different low temperatures.

## 3. Discussion

### 3.1 Analysis of experimental results

The modulation in the amplitude and phase of the transmission originates from the change in conductivity with temperature. As the conductivity increases with reducing temperature, the magnitude of circular currents at the LC resonance and the linear currents at the dipole resonance increases, resulting in strengthening of both resonances [29-31]. According to the two-fluid model, the real part of the conductivity ($\sigma_r$) of YBCO is contributed by the normal state carriers whose motion follows Drude Model while the imaginary part ($\sigma_i$) is determined by both the normal state carries and the superconducting carries, which follow the London Equation [32]. At terahertz frequencies, for the temperatures higher than $T_c$ = 86 K the real part of conductivity dominates since the absolute value of imaginary conductivity is three orders of magnitude less than the real part. But starting from several degrees under $T_c$ the imaginary conductivity value rises sharply with falling temperatures and the total conductivity is dominated by $\sigma_i$. We measured the terahertz direct transmission through unpatterned 280nm thick YBCO film and extracted the real part of conductivity at 297, 200, 150, and at 85 K. No obvious frequency dependence is observed but as the temperature goes down, $\sigma_r$ increases from $3.6 \times 10^5$ S/m at 297 K to $7.1 \times 10^5$ S/m at 85 K. Because the dissipation loss is inversely proportional to $\sigma_r$ it is expected that the dip in the transmission amplitude goes deeper as the temperature reduces and the resonance features get stronger.

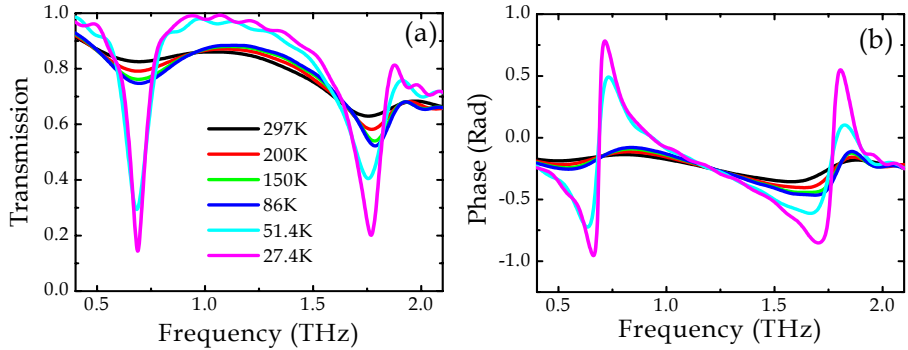

Fig. 3. (color online). Simulated (a) amplitude and (b) phase transmission spectra of the YBCO MM above and below the transition temperature.

### 3.2 Simulation and analysis above and below $T_c$

Previous works on conductivity of YBCO film at terahertz frequencies have shown that below $T_c$ [33,34] $\sigma_i$ increases dramatically and exceeds the real part at several Kelvins below $T_c$. The sharp rise in the imaginary conductivity is the signature of the onset of superconductivity and it can serve as an independent method for the determination of superconducting transition temperature. In our measurements the YBCO does not reach its superconducting stage, therefore the MM resonance is not extremely sharp. Using the measured real conductivity at 297, 200, 150, 85 K and the complex conductivity from literature at 51.5 and 27.4 K, we carried out the CST simulation of the superconductor MM to investigate the nature of resonances at temperatures above and well below the transition temperature [33]. Figure 3a reveals the dramatic change in the transmission at the LC resonance as the temperature reaches well below $T_c$ at 51.4 and 27.4 K. The amplitude dip at 0.75 THz is as low as 0.23 and the Q-factor reaches 6.75 which are normal in the terahertz regime [35]. Besides this the sharp phase change as large as 1.7 radians is attained around LC resonance which is high in terahertz MM shown in Fig. 3b. The real and imaginary parts of the conductivity at 27.4 K are ~$4.3 \times 10^5$ S/m and ~$2.3 \times 10^6$ S/m respectively at 0.75 THz [33], which gives a skin depth of $\delta = 330$ nm and a London Penetration Depth $\lambda_L < 186$ nm. Since our YBCO MM metamaterial film has a thickness higher than $\lambda_L$, the current in SRR grows stronger due to the superconducting carriers and it results in a sharp inductor-capacitor resonance. It should be stressed here that the absolute ratio of real to imaginary conductivity decreases sharply below the superconducting temperature and this in turn supports higher circular currents in the LC resonator and thus a stronger resonance. This behavior has also been observed in regular thin film SRRs made up of Drude metals at terahertz frequencies [36].

Although the use of superconductor films is one of the paths to minimize the losses in MMs, the quality of YBCO film used depends highly on dopant species, growth conditions, the composition, process optimization and substrate preparation. The critical current density in YBCO is greatly limited by material imperfections such as cracks, voids, or secondary phases that partially block the current flow. Another factor limiting the current flow is the increasing misorientation angle between the adjacent grains. These defects lead to the losses in the superconductor MM and prevent the SRRs from supporting extremely high Q resonances. The YBCO film does not seem to reach its superconducting temperature in our measurement at 85 K and this could be due to several defect parameters involved. However, the simulations at 51.4 and 27.4 K clearly reveal a switching effect in the SRR behavior below the transition temperature.

## 4. Summary

In conclusion, we demonstrate the temperature tunable behavior of high temperature superconductor MM at terahertz frequencies. The kinetic inductance of the superconductor film is sensitive to the surrounding temperature and thus allows the MM response to be altered with the reducing temperature. The amplitude transmission shows gradual decrease at the LC and dipole resonances as the MM is cooled up to 85 K but the resonance response changes dramatically below the transition temperature of YBCO. Such MMs can find potential applications in designing low-loss temperature controlled terahertz devices and components.

**Acknowledgements**

The authors thank J. Wu, J. Zhang, and S. Zhang for their support and discussions. This work was supported by the U.S. National Science Foundation (Grant No. ECCS-0725764), the China Scholarship Council, the National Key Basic Research Special Foundation of China (Grant Nos. 2007CB310403 and 2007CB310408), the Tianjin Sci-Tech Support Program (Grant No. 8ZCKFZC28000), and the Doctoral Program for Higher Education of China (Grant No. 200800560026). J. Han thanks financial support from the MOE Academic Research Fund of Singapore and the Lee Kuan Yew Fund.